\documentclass[aps,prd,preprintnumbers,showpacs,twocolumn,groupedaddress,nofootinbib]{revtex4}
\usepackage{graphicx}
\usepackage{latexsym}
\def\beq{\begin{equation}}
\def\eel{\end{equation}}
\def\bey{\begin{eqnarray}}
\def\eey{\end{eqnarray}}

\def\lsim{\mathrel{\raise.3ex\hbox{$<$\kern-.75em\lower1ex\hbox{$\sim$}}}}
\def\gsim{\mathrel{\raise.3ex\hbox{$>$\kern-.75em\lower1ex\hbox{$\sim$}}}}

\begin{document}

\title{Is the CMB telling us that dark matter is weaker than weakly interacting?}  
\author{Dan Hooper}
\affiliation{Center for Particle Astrophysics, Fermi National Accelerator Laboratory, Batavia, IL 60510, USA}
\affiliation{Department of Astronomy and Astrophysics, University of  Chicago, Chicago, IL 60637, USA}

\date{\today}

\begin{abstract}

If moduli, or other long-lived heavy states, decay in the early universe in part into light and feebly interacting particles (such as axions), these decay products could account for the additional energy density in radiation that is suggested by recent measurements of the CMB. These moduli decays will also, however, alter the expansion history of the early universe, potentially diluting the thermal relic abundance of dark matter. If this is the case, then dark matter particles must annihilate with an even lower cross section than required in the standard thermal scenario ($\langle \sigma v\rangle \ll 3\times 10^{-26}$ cm$^3$/s) if they are to make up the observed density of dark matter. This possibility has significant implications for direct and indirect searches for dark matter.

\end{abstract}

\pacs{95.35.+d, 98.80.Cq, 98.80.-k; FERMILAB-PUB-13-244-A}
\maketitle

Among other cosmological parameters, the patterns of the temperature fluctuations in the cosmic microwave background (CMB) can be used to determine the energy density of relativistic particles in the early universe. This is typically described in terms of the effective number of neutrino species, $N_{\rm eff}$. Although the standard model with three light neutrino species predicts a value of $N_{\rm eff} = 3.046$, measurements of the CMB~\cite{planck,wmap,spt,act}, when combined with those of the Hubble constant~\cite{hst,Freedman:2012ny} and baryon acoustic oscillations in the matter power spectrum~\cite{Percival:2009xn,Padmanabhan:2012hf,Blake:2011en,Anderson:2012sa,Beutler:2011hx}, appear to favor a somewhat higher value: 
$N_{\rm eff}=3.52^{+0.48}_{-0.45}$ (at the 95\% confidence level)~\cite{planck}. If these indications of $N_{\rm eff} > 3.046$ are supported by future measurements, it would imply that the early universe contained a higher energy density in relativistic particles than can be accounted for in the standard model. Exotic particle species that could potentially account for this energy density is sometimes referred to as {\it dark radiation}. 

A natural  way to generate dark radiation is through the decays of long-lived massive particles~\cite{Ichikawa:2007jv,Hasenkamp:2011em,Menestrina:2011mz,Kobayashi:2011hp,Jeong:2012np,Fischler:2010xz,Higaki:2012ba,Choi:2012zna,Graf:2012hb,Cicoli:2012aq,Higaki:2012ar,Bae:2013qr,Jeong:2013axf,Graf:2013xpe,Hooper:2011aj,Kelso:2013paa,GonzalezGarcia:2012yq,Conlon:2013isa,Conlon:2013txa,Angus:2013zfa} (for a review, see Ref.~\cite{Hasenkamp:2012ii}). A particularly well motivated example of such particles are moduli, which are generically expected within the context of string theory, with various theoretical arguments favoring a mass range of $m_{\Phi} \sim 10^4-10^7$ GeV~\cite{Moroi:1999zb,de Carlos:1993jw,Choi:2005ge,Acharya:2008bk,Blumenhagen:2009gk} (moduli lighter than tens of TeV are in conflict with the successful predictions of big bang nucleosynthesis~\cite{Coughlan:1983ci,Ellis:1986zt,de Carlos:1993jw,Banks:1993en}). Although we will limit our discussion to the case of decaying moduli, our conclusions apply to any very heavy, unstable state with highly suppressed interactions. 

\begin{figure*}[t!]
\includegraphics[width=3.43in]{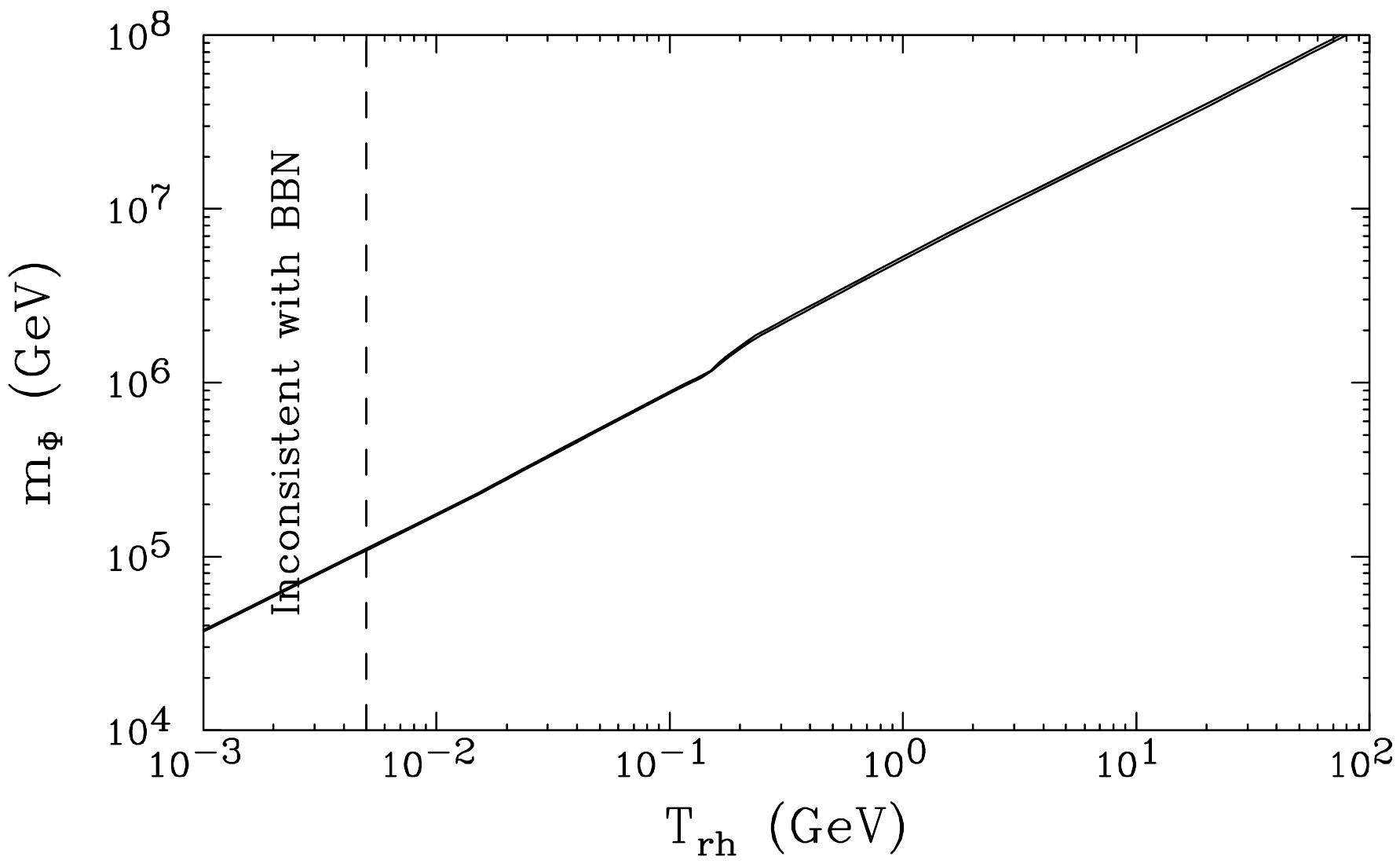}
\hspace{0.3cm}
\includegraphics[width=3.43in]{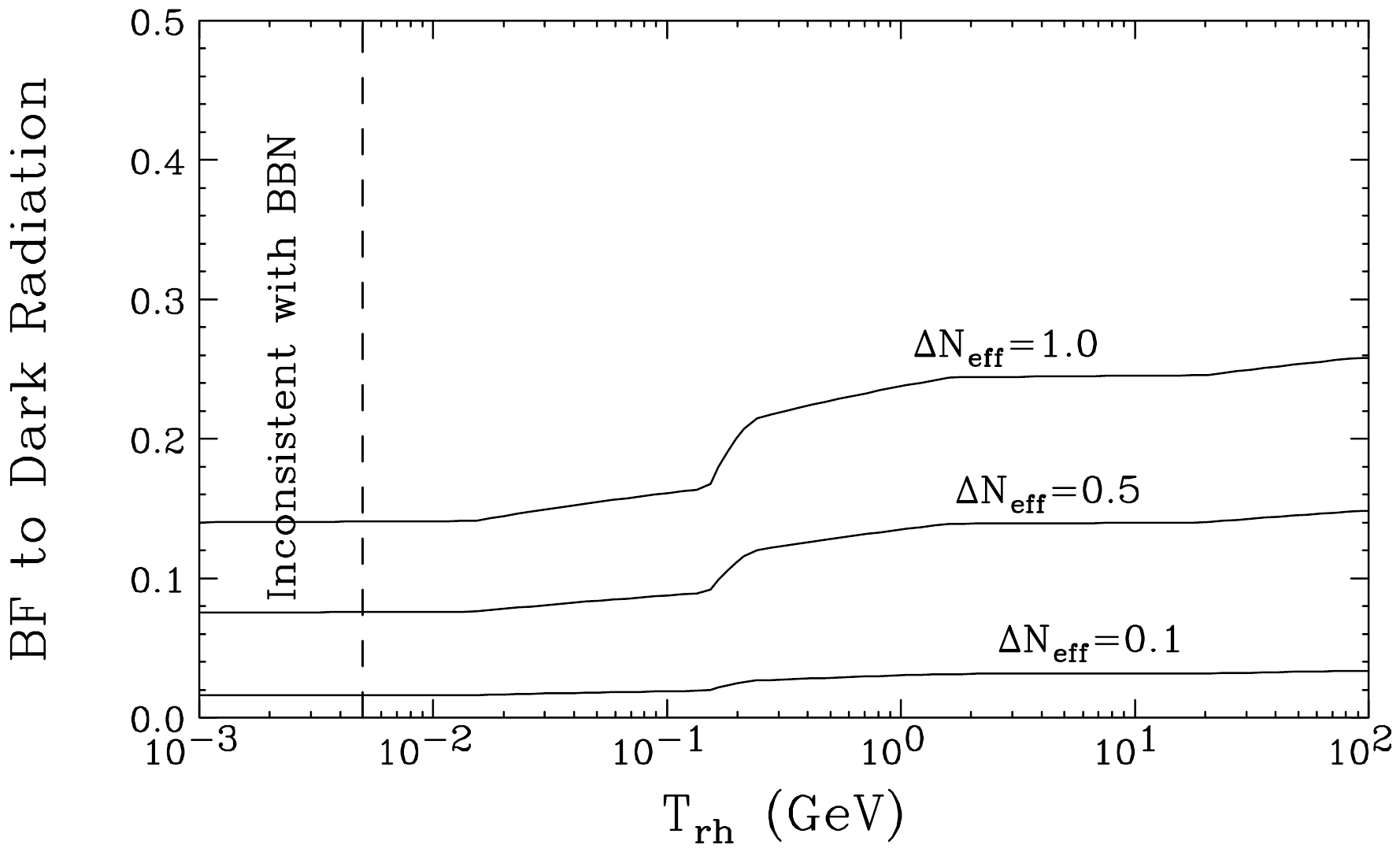}
\caption{Left frame: The temperature to which moduli decays reheat the universe, as a function of moduli mass, for the case of $\kappa=1$. Right frame: The contribution from moduli decay to the effective number of neutrino species, $\Delta N_{\rm eff}$, as a function of the reheating temperature and the branching fraction for moduli decays into axions or other dark radiation. For a wide range of reheating temperatures (and moduli masses), branching fractions on the order of 10\% for moduli to decay to dark radiation can accommodate the value favored by recent CMB data, $\Delta N_{\rm eff} =  0.474^{+0.48}_{-0.45}$ (95\% CL).} 
\label{reheat}
\end{figure*}

If moduli exist, they will naturally come to dominate the universe's energy density (see, for example, Refs.~\cite{Banks:1993en,de Carlos:1993jw,Coughlan:1983ci}) and will reheat the universe upon their decay. Moduli, or other states with Planck-scale suppressed couplings, decay at a rate given by:
\begin{equation}
\Gamma = \frac{1}{4} \frac{m^3_{\Phi}}{(M_{\rm Pl}/\kappa)^2},
\end{equation}
where $\kappa$ is a model dependent order one factor, and $M_{\rm Pl}=2.435 \times 10^{18}$ GeV is the reduced Planck mass. This corresponds to a lifetime of $\tau \simeq 0.05 \,\, {\rm sec} \, \times (1/\kappa)^2 \, (100\,{\rm TeV}/m_{\Phi})^3$. The decays of these moduli produce relativistic particles and entropy, reheating the universe to a temperature given by:
\begin{equation}
T_{\rm rh} = \kappa \bigg(\frac{5 \, B_{\rm SM}}{2 \pi^4 \, g_{\star}(T_{\rm rh})}\bigg)^{1/4} \, \bigg(\frac{m^3_{\Phi}}{M_{Pl}}\bigg)^{1/2},
\end{equation}
where $B_{\rm SM}$ is the fraction of moduli decays that produce standard model particles, and $g_{\star}(T_{\rm rh})$ is the number of degrees-of-freedom evaluated at the temperature of reheating. This relationship is shown in the left frame of Fig.~\ref{reheat}, for $\kappa=1$. If the moduli decay not only to standard model particles, but also a fraction of the time, $B_a = 1-B_{\rm SM}$, to very light and feebly interacting particles (such as axions, hidden sector photons, etc.), those particles will contribute to the energy density in radiation in the early universe, altering the observed value of $N_{\rm eff}$~\cite{Conlon:2013isa,Higaki:2013lra}:
\begin{equation}
\Delta N_{\rm eff} = \frac{43}{7} \, \frac{B_{a}}{1-B_a} \, \bigg(\frac{g_{\star}(T_{\nu \, decoupling})}{g_{\star}(T_{\rm rh})}\bigg)^{1/3}, 
\end{equation}
where $g_{\star} \simeq 10.75$ at the temperature of neutrino decoupling. 

In the right frame of Fig.~\ref{reheat}, we show the contribution to $N_{\rm eff}$ predicted in this scenario, as a function of the reheating temperature and the branching fraction for moduli decay to axions (or other dark radiation). We find that for a very wide range of reheating temperatures, branching fractions to axions in the range of $\sim2$-25\% yield contributions comparable to that suggested by CMB observations. Interestingly, moduli branching fractions to axions are naturally expected to be of this order, $B_a = \mathcal O(0.1)$~\cite{Higaki:2013lra}.

The late time decays of moduli or other heavy states can also have a very significant impact on the process of thermal freeze-out of dark matter and on the resulting relic density produced in the early universe~\cite{Chung:1998rq,Fornengo:2002db,Pallis:2004yy,Gelmini:2006pw,Gelmini:2006pq,Allahverdi:2013tca} (for a review, see Ref.~\cite{Watson:2009hw}). In the standard thermal scenario, weakly interacting dark matter particles freeze-out in a highly radiation dominated universe, at the temperature at which the effects of Hubble expansion first overcome those of annihilation ($T_{\rm FO}$). If there exist, however, a long-lived massive particle species with couplings too feeble to be maintained in thermal equilibrium ({\it e.g.} moduli), they will naturally come to dominate the energy density of the universe (this can be seen by simply comparing the scalings of the densities of relativistic and non-relativistic particles, $\rho \propto a^{-4}$ and $\rho \propto a^{-3}$, respectively). If these decays reheat the universe to a temperature that is below that of the dark matter's freeze-out temperature, then the expansion rate of the universe will be faster during the period of time between freeze-out and reheating, diluting the density of dark matter by a factor of $(T_{\rm rh}/T_{\rm FO})^3$.

For illustration, consider dark matter within a simple supersymmetric scenario: a 200 GeV bino-like neutralino dark matter candidate, 2 TeV squarks and gluinos (to evade LHC constraints~\cite{squarksgluinos}), 700 GeV for all other mass parameters, $\tan \beta=10$, and a value of the top trilinear coupling chosen to accommodate the observed mass of the light Higgs. If it were not for cosmological constraints, such a model would be perfectly reasonable. But, assuming the standard thermal history, the neutralino relic abundance predicted in this model is $\Omega_{\chi} h^2 \approx 5.1$~\cite{micromegas}, well above the measured value of 0.12. And although we can bring this model into line with the observed cold dark matter density by simply lowering the higgsino mass to $\mu \simeq 240$ GeV, leading to a mixed bino-higgsino neutralino with sufficient couplings to avoid being overproduced in the early universe, these same couplings also increase the dark matter's spin-independent elastic scattering cross section with nuclei to $3.4\times 10^{-44}$ cm$^2$, which is almost an order of magnitude larger than the upper limit placed by the XENON-100 collaboration~\cite{xenon}.

The above example illustrates two key points. First, while the ``WIMP miracle'' has provided a great deal of motivation for dark matter candidates with weak-scale masses and interactions, in many theoretical frameworks (including minimal supersymmetry) the connection between electroweak-scale physics and the measured dark matter abundance is very tenuous. The vast majority of the otherwise viable supersymmetric parameter space predicts a thermal relic abundance of neutralinos that is in considerable excess of the observed dark matter density.  Second, if one naively increases the couplings of their dark matter candidate to avoid being thermally overproduced in the early universe, this also tends to increase the dark matter's elastic scattering cross section with nuclei to values above current experimental constraints. Only in rather special regions of parameter space, such as those in which the lightest neutralino efficiently co-annihilates with another nearly-degenerate sparticle (such as a stau, chargino, or stop)~\cite{Griest:1990kh} or annihilates through a somewhat fine-tuned resonance~\cite{Griest:1990kh,Hooper:2013qjx,Han:2013gba}, can the observed dark matter abundance match the predicted thermal relic density without exceeding the constraints placed by direct detection experiments.\footnote{There also exist regions of supersymmetric parameter space in which a mixed gaugino-higgsino can satisfying these constraints~\cite{Buchmueller:2012hv,ArkaniHamed:2006mb,Feng:2011aa}, although direct detection experiments should be able to exclude such models in the near future if no signal is detected~\cite{lux,xenon1t}.} Note that these arguments are not limited to the specific case of neutralino dark matter, but apply to a broad range of weakly interacting dark matter candidates (for example, see Ref.~\cite{Beltran:2008xg}).

\begin{figure}[t!]
\includegraphics[width=3.43in]{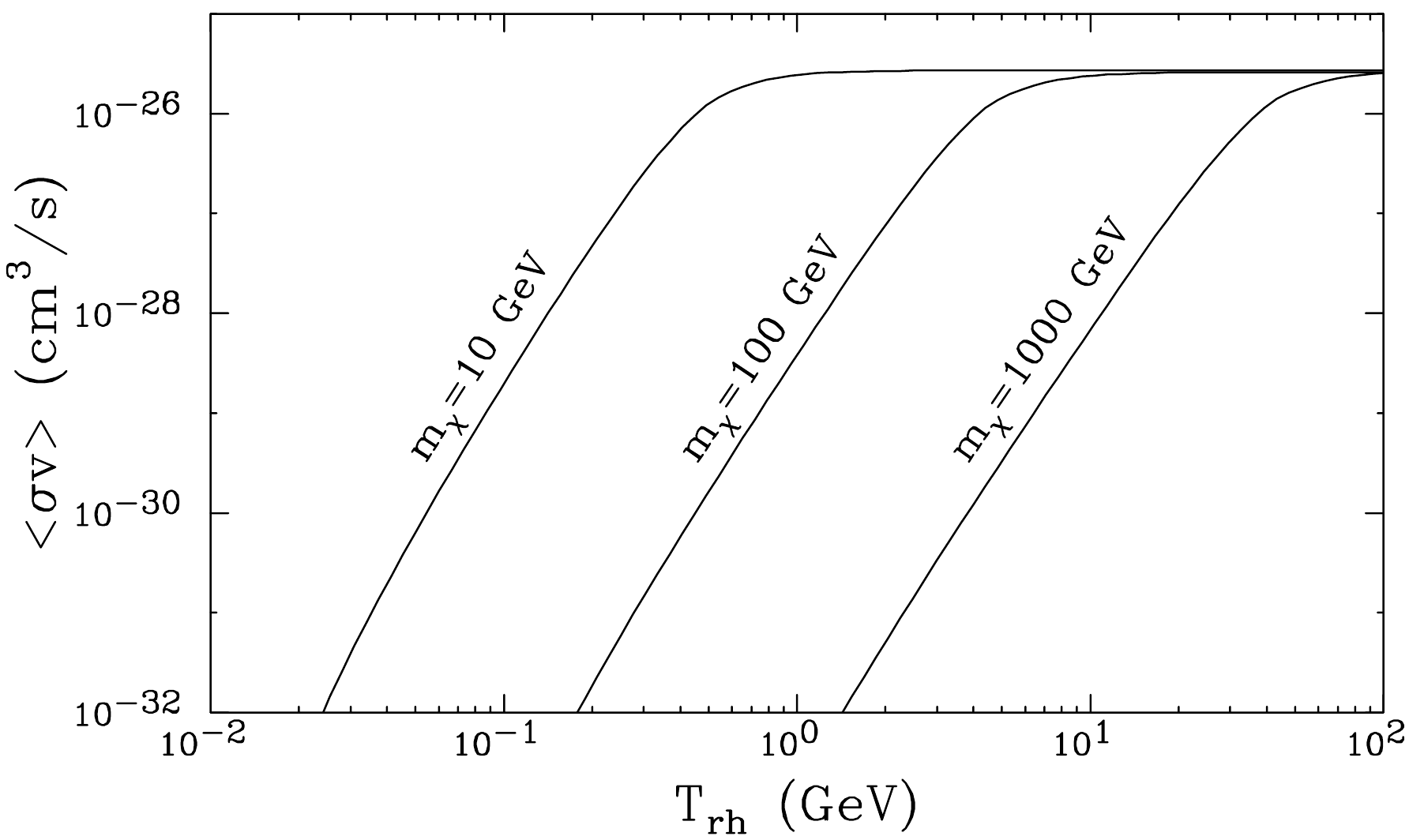}
\caption{The annihilation cross section, thermally averaged at the temperature of freeze-out, required for a particle to generate the observed cosmological dark matter abundance, as a function of the reheating temperature. For high reheating temperatures, $T_{\rm rm} > T_{\rm FO}$, we recover the standard value expected for a thermal relic.  For lower reheating temperatures, viable dark matter candidates must annihilate with smaller cross sections than are predicted in the standard scenario.} 
\label{relic}
\end{figure}

We reach a very different conclusion, however, if we consider the same supersymmetry model within the context of a scenario with a low reheating temperature induced by moduli decay.  A 200 GeV neutralino will undergo thermal freeze-out at a temperature of approximately $T_{\rm FO} \approx m_{\chi}/20 \approx 10$ GeV. From Fig.~\ref{reheat}, we see than for moduli lighter than $\sim$20 PeV, the decays will reheat the universe to a temperature below the freeze-out temperature, and thus will significantly dilute the abundance of neutralinos. Considering $m_{\Phi}=10$~PeV and $\kappa=1$, for example, the moduli decays reheat the universe to a temperature of approximately 2.8 GeV, diluting the neutralino dark matter density by a factor of (2.8 GeV/10 GeV)$^3\sim 0.02$.\footnote{The change in the temperature of freeze-out, which now occurs during matter domination, also has a small effect on the results of this calculation. We include this effect explicitly in the contours shown in Fig.~\ref{relic}.} For the same supersymmetric parameters described two paragraphs above, and for $\mu \approx 1$ TeV, this degree of dilution yields a neutralino abundance that is consistent with the measured density of dark matter, and predicts a spin-independent elastic scattering cross section of $\sigma_{\rm SI} \approx 2\times 10^{-46}$ cm$^2$, well below current constraints.

In Fig.~\ref{relic}, we show the annihilation cross section (thermally averaged at the temperature of freeze-out) required for a 10 GeV, 100 GeV or 1000 GeV particle to make up the observed dark matter abundance, as a function of the reheating temperature. For reheating temperatures greater than the freeze-out temperature, we recover the standard value for a thermal relic, $\langle \sigma v \rangle \simeq (2-3) \times 10^{-26}$ cm$^3$/s~\cite{Steigman:2012nb}. For lower reheating temperatures (corresponding to lower moduli masses), however, lower annihilation cross sections are required; in some cases dramatically lower.  If the couplings and exchanged particles responsible for the dark matter's annihilation cross section are also those (or are related to those) which generate its elastic scattering cross section nuclei, we would expect rates at direct detection experiments to be suppressed by a similar factor.

Throughout this letter, we have assumed that the dark matter abundance is dominated by thermal relics (albeit with a density diluted by late time reheating). Moduli may also produce dark matter particles directly, as non-thermal decay products. In such scenarios, the resulting dark matter will have an annihilation cross section that is {\it larger} than predicted in the standard thermal scenario (parametrically, larger by a factor of $\sim$$T_{\rm FO}/T_{\rm rh}$~\cite{Watson:2009hw}). A well studied example is dark matter in the form of non-thermally produced winos~\cite{Moroi:1999zb,Acharya:2008bk}. The large annihilation cross sections predicted for dark matter originating as moduli decay products lead to strong constraints from indirect detection experiments; gamma-ray observations presently limit wino dark matter to $m_{\tilde{W}} \gsim 700$ GeV, for example~\cite{GeringerSameth:2011iw,Ackermann:2011wa,Hooper:2012sr}. The combination of such indirect detection constraints and the increasingly stringent limits from direct detection experiments appear to disfavor dark matter that is produced through non-thermal decays, although some scenarios remain viable. Fortunately, it is not difficult to suppress the branching fractions of moduli to superpartners (or, more generally, to the sector carrying the parity or charge that stabilizes the dark matter). For example, despite the fact that supersymmetry requires moduli to have identical couplings to gauge bosons and gauginos, their decay to gauginos is chirality suppressed by a factor of $(m_{\chi}/m_{\Phi})^2$. For weak-scale gaugino masses, this naturally leads to very small branching fractions to supersymmetric particles, on the order of $\sim 10^{-8} \times (1 \, {\rm PeV}/m_{\Phi})^2$~\cite{Dine:1995kz,Moroi:1999zb}.

At this time, we will summarize the main points made in this letter:
\begin{itemize}
\item{If moduli (or other long-lived heavy states) decay with an order 10\% branching fraction to axions (or other very light and feebly interacting particles), this could account for the ``dark radiation'' suggested by recent CMB observations, $N_{\rm eff}=3.52^{+0.48}_{-0.45}$(95\% CL) $>$ 3.046.}
\item{If the moduli have a mass in the range of tens of TeV to tens of PeV, they will alter the expansion history of the universe, resulting in the dilution of the thermal abundance of dark matter particles.}
\item{When this dilution is taken into account, we find that the dark matter's annihilation cross section must be smaller, and possibly much smaller, than predicted in the standard thermal scenario.}
\item{If the particles exchanged in the process of dark matter annihilation are also those responsible for elastic scattering with nuclei, then we also expect lower rates in direct detection experiments than would be predicted in the standard thermal scenario.}
\item{Whereas supersymmetric dark matter with a standard thermal history predicts a thermal abundance of neutralinos that is in considerable excess of the observed dark matter density in all but a few corners of parameter space (such as the focus point, co-annihilation, and resonance regions), the dilution of the thermal abundance via moduli decay opens up a large range of less finely tuned models.} 
\end{itemize}

In other words, moduli decay provides a simple and well-motivated way to explain the high value of $\Delta N_{\rm eff}$ suggested by observations, and such decays will also alter the thermal history of the early universe. Once this is taken into account, we find that dark matter particles may annihilate with a much smaller cross section than is predicted for a standard thermal history -- the dark matter may interact more weakly than weak.  The prospects for direct and indirect detection may be suppressed in such scenarios, relieving the tension introduced by the null results of XENON100 and other experiments searching for dark matter.

\bigskip
%\newpage

{\it Acknowledgements}:  The author would like to thank Scott Watson, Lisa Randall, and Graciela Gelmini for valuable discussions, and the Kavli Institute for Theoretical Physics for their hospitality. This work has been supported by the US Department of Energy.


\begin{thebibliography}{}

  
  \bibitem{planck} 
  P.~A.~R.~Ade {\it et al.}  [Planck Collaboration],
  %``Planck 2013 results. XVI. Cosmological parameters,''
  arXiv:1303.5076 [astro-ph.CO].
  %%CITATION = ARXIV:1303.5076;%%
  %290 citations counted in INSPIRE as of 21 Jun 2013
  
  \bibitem{wmap} 
  G.~Hinshaw {\it et al.}  [WMAP Collaboration],
  %``Nine-Year Wilkinson Microwave Anisotropy Probe (WMAP) Observations: Cosmological Parameter Results,''
  arXiv:1212.5226 [astro-ph.CO].
  %%CITATION = ARXIV:1212.5226;%%
  %315 citations counted in INSPIRE as of 21 Jun 2013
  
  \bibitem{spt} 
  Z.~Hou, C.~L.~Reichardt, K.~T.~Story, B.~Follin, R.~Keisler, K.~A.~Aird, B.~A.~Benson and L.~E.~Bleem {\it et al.},
  %``Constraints on Cosmology from the Cosmic Microwave Background Power Spectrum of the 2500-square degree SPT-SZ Survey,''
  arXiv:1212.6267 [astro-ph.CO].
  %%CITATION = ARXIV:1212.6267;%%
  %69 citations counted in INSPIRE as of 21 Jun 2013
  
  \bibitem{act} 
  J.~L.~Sievers, R.~A.~Hlozek, M.~R.~Nolta, V.~Acquaviva, G.~E.~Addison, P.~A.~R.~Ade, P.~Aguirre and M.~Amiri {\it et al.},
  %``The Atacama Cosmology Telescope: Cosmological parameters from three seasons of data,''
  arXiv:1301.0824 [astro-ph.CO].
  %%CITATION = ARXIV:1301.0824;%%
  %65 citations counted in INSPIRE as of 21 Jun 2013
  
  \bibitem{hst} 
  A.~G.~Riess, L.~Macri, S.~Casertano, H.~Lampeitl, H.~C.~Ferguson, A.~V.~Filippenko, S.~W.~Jha and W.~Li {\it et al.},
  %``A 3% Solution: Determination of the Hubble Constant with the Hubble Space Telescope and Wide Field Camera 3,''
  Astrophys.\ J.\  {\bf 730}, 119 (2011)
  [Erratum-ibid.\  {\bf 732}, 129 (2011)]
  [arXiv:1103.2976 [astro-ph.CO]].
  %%CITATION = ARXIV:1103.2976;%%
  %317 citations counted in INSPIRE as of 21 Jun 2013
  
  \bibitem{Freedman:2012ny} 
  W.~L.~Freedman, B.~F.~Madore, V.~Scowcroft, C.~Burns, A.~Monson, S.~E.~Persson, M.~Seibert and J.~Rigby,
  %``Carnegie Hubble Program: A Mid-Infrared Calibration of the Hubble Constant,''
  Astrophys.\ J.\  {\bf 758}, 24 (2012)
  [arXiv:1208.3281 [astro-ph.CO]].
  %%CITATION = ARXIV:1208.3281;%%
  %35 citations counted in INSPIRE as of 21 Jun 2013

  
\bibitem{Percival:2009xn} 
  W.~J.~Percival {\it et al.}  [SDSS Collaboration],
  %``Baryon Acoustic Oscillations in the Sloan Digital Sky Survey Data Release 7 Galaxy Sample,''
  Mon.\ Not.\ Roy.\ Astron.\ Soc.\  {\bf 401}, 2148 (2010)
  [arXiv:0907.1660 [astro-ph.CO]].
  %%CITATION = ARXIV:0907.1660;%%
  %691 citations counted in INSPIRE as of 21 Jun 2013  
  
  \bibitem{Padmanabhan:2012hf} 
  N.~Padmanabhan, X.~Xu, D.~J.~Eisenstein, R.~Scalzo, A.~J.~Cuesta, K.~T.~Mehta and E.~Kazin,
  %``A 2 per cent distance to $z$=0.35 by reconstructing baryon acoustic oscillations - I. Methods and application to the Sloan Digital Sky Survey,''
  Mon.\ Not.\ Roy.\ Astron.\ Soc.\  {\bf 427}, no. 3, 2132 (2012)
  [arXiv:1202.0090 [astro-ph.CO]].
  %%CITATION = ARXIV:1202.0090;%%
  %68 citations counted in INSPIRE as of 21 Jun 2013
  
  \bibitem{Blake:2011en} 
  C.~Blake, E.~Kazin, F.~Beutler, T.~Davis, D.~Parkinson, S.~Brough, M.~Colless and C.~Contreras {\it et al.},
  %``The WiggleZ Dark Energy Survey: mapping the distance-redshift relation with baryon acoustic oscillations,''
  Mon.\ Not.\ Roy.\ Astron.\ Soc.\  {\bf 418}, 1707 (2011)
  [arXiv:1108.2635 [astro-ph.CO]].
  %%CITATION = ARXIV:1108.2635;%%
  %136 citations counted in INSPIRE as of 21 Jun 2013
  
  \bibitem{Anderson:2012sa} 
  L.~Anderson, E.~Aubourg, S.~Bailey, D.~Bizyaev, M.~Blanton, A.~S.~Bolton, J.~Brinkmann and J.~R.~Brownstein {\it et al.},
  %``The clustering of galaxies in the SDSS-III Baryon Oscillation Spectroscopic Survey: Baryon Acoustic Oscillations in the Data Release 9 Spectroscopic Galaxy Sample,''
  Mon.\ Not.\ Roy.\ Astron.\ Soc.\  {\bf 427}, no. 4, 3435 (2013)
  [arXiv:1203.6594 [astro-ph.CO]].
  %%CITATION = ARXIV:1203.6594;%%
  %119 citations counted in INSPIRE as of 21 Jun 2013
  
  \bibitem{Beutler:2011hx} 
  F.~Beutler, C.~Blake, M.~Colless, D.~H.~Jones, L.~Staveley-Smith, L.~Campbell, Q.~Parker and W.~Saunders {\it et al.},
  %``The 6dF Galaxy Survey: Baryon Acoustic Oscillations and the Local Hubble Constant,''
  Mon.\ Not.\ Roy.\ Astron.\ Soc.\  {\bf 416}, 3017 (2011)
  [arXiv:1106.3366 [astro-ph.CO]].
  %%CITATION = ARXIV:1106.3366;%%
  %128 citations counted in INSPIRE as of 21 Jun 2013
  
 %%%%%%
%9
\bibitem{Ichikawa:2007jv} 
  K.~Ichikawa, M.~Kawasaki, K.~Nakayama, M.~Senami and F.~Takahashi,
  %``Increasing effective number of neutrinos by decaying particles,''
  JCAP {\bf 0705}, 008 (2007)
  [hep-ph/0703034 [HEP-PH]].
  %%CITATION = HEP-PH/0703034;%%
  %57 citations counted in INSPIRE as of 21 Jun 2013


\bibitem{Fischler:2010xz} 
  W.~Fischler and J.~Meyers,
  %``Dark Radiation Emerging After Big Bang Nucleosynthesis?,''
  Phys.\ Rev.\ D {\bf 83}, 063520 (2011)
  [arXiv:1011.3501 [astro-ph.CO]].
  %%CITATION = ARXIV:1011.3501;%%
  %35 citations counted in INSPIRE as of 21 Jun 2013

  
\bibitem{Hasenkamp:2011em} 
  J.~Hasenkamp,
  %``Dark radiation from the axino solution of the gravitino problem,''
  Phys.\ Lett.\ B {\bf 707}, 121 (2012)
  [arXiv:1107.4319 [hep-ph]].
  %%CITATION = ARXIV:1107.4319;%%
  %27 citations counted in INSPIRE as of 21 Jun 2013

\bibitem{Menestrina:2011mz} 
  J.~L.~Menestrina and R.~J.~Scherrer,
  %``Dark Radiation from Particle Decays during Big Bang Nucleosynthesis,''
  Phys.\ Rev.\ D {\bf 85}, 047301 (2012)
  [arXiv:1111.0605 [astro-ph.CO]].
  %%CITATION = ARXIV:1111.0605;%%
  %27 citations counted in INSPIRE as of 21 Jun 2013

\bibitem{Kobayashi:2011hp} 
  T.~Kobayashi, F.~Takahashi, T.~Takahashi and M.~Yamaguchi,
  %``Dark Radiation from Modulated Reheating,''
  JCAP {\bf 1203}, 036 (2012)
  [arXiv:1111.1336 [astro-ph.CO]].
  %%CITATION = ARXIV:1111.1336;%%
  %14 citations counted in INSPIRE as of 21 Jun 2013

\bibitem{Hooper:2011aj} 
  D.~Hooper, F.~S.~Queiroz and N.~Y.~Gnedin,
  %``Non-Thermal Dark Matter Mimicking An Additional Neutrino Species In The Early Universe,''
  Phys.\ Rev.\ D {\bf 85}, 063513 (2012)
  [arXiv:1111.6599 [astro-ph.CO]].
  %%CITATION = ARXIV:1111.6599;%%
  %25 citations counted in INSPIRE as of 21 Jun 2013

\bibitem{Jeong:2012np} 
  K.~S.~Jeong and F.~Takahashi,
  %``Light Higgsino from Axion Dark Radiation,''
  JHEP {\bf 1208}, 017 (2012)
  [arXiv:1201.4816 [hep-ph]].
  %%CITATION = ARXIV:1201.4816;%%
  %17 citations counted in INSPIRE as of 21 Jun 2013

\bibitem{Higaki:2012ba} 
  T.~Higaki, K.~Kamada and F.~Takahashi,
  %``Higgs, Moduli Problem, Baryogenesis and Large Volume Compactifications,''
  JHEP {\bf 1209}, 043 (2012)
  [arXiv:1207.2771 [hep-ph]].
  %%CITATION = ARXIV:1207.2771;%%
  %11 citations counted in INSPIRE as of 21 Jun 2013

\bibitem{Choi:2012zna} 
  K.~Choi, K.~-Y.~Choi and C.~S.~Shin,
  %``Dark radiation and small-scale structure problems with decaying particles,''
  Phys.\ Rev.\ D {\bf 86}, 083529 (2012)
  [arXiv:1208.2496 [hep-ph]].
  %%CITATION = ARXIV:1208.2496;%%
  %18 citations counted in INSPIRE as of 21 Jun 2013
  
\bibitem{Graf:2012hb} 
  P.~Graf and F.~D.~Steffen,
  %``Axions and saxions from the primordial supersymmetric plasma and extra radiation signatures,''
  JCAP {\bf 1302}, 018 (2013)
  [arXiv:1208.2951 [hep-ph]].
  %%CITATION = ARXIV:1208.2951;%%
  %14 citations counted in INSPIRE as of 21 Jun 2013  

\bibitem{Cicoli:2012aq} 
  M.~Cicoli, J.~P.~Conlon and F.~Quevedo,
  %``Dark Radiation in LARGE Volume Models,''
  Phys.\ Rev.\ D {\bf 87}, 043520 (2013)
  [arXiv:1208.3562 [hep-ph]].
  %%CITATION = ARXIV:1208.3562;%%
  %25 citations counted in INSPIRE as of 21 Jun 2013

\bibitem{Higaki:2012ar} 
  T.~Higaki and F.~Takahashi,
  %``Dark Radiation and Dark Matter in Large Volume Compactifications,''
  JHEP {\bf 1211}, 125 (2012)
  [arXiv:1208.3563 [hep-ph]].
  %%CITATION = ARXIV:1208.3563;%%
  %20 citations counted in INSPIRE as of 21 Jun 2013

  \bibitem{GonzalezGarcia:2012yq} 
  M.~C.~Gonzalez-Garcia, V.~Niro and J.~Salvado,
  %``Dark Radiation and Decaying Matter,''
  JHEP {\bf 1304}, 052 (2013)
  [arXiv:1212.1472 [hep-ph]].
  %%CITATION = ARXIV:1212.1472;%%
  %9 citations counted in INSPIRE as of 21 Jun 2013


\bibitem{Bae:2013qr} 
  K.~J.~Bae, H.~Baer and A.~Lessa,
  %``Dark Radiation Constraints on Mixed Axion/Neutralino Dark Matter,''
  JCAP {\bf 1304}, 041 (2013)
  [arXiv:1301.7428 [hep-ph]].
  %%CITATION = ARXIV:1301.7428;%%
  %12 citations counted in INSPIRE as of 21 Jun 2013

\bibitem{Jeong:2013axf} 
  K.~S.~Jeong and F.~Takahashi,
  %``Axionic Co-genesis of Baryon, Dark Matter and Dark Radiation,''
  JHEP {\bf 1304}, 121 (2013)
  [arXiv:1302.1486 [hep-ph]].
  %%CITATION = ARXIV:1302.1486;%%
  %6 citations counted in INSPIRE as of 21 Jun 2013

\bibitem{Graf:2013xpe} 
  P.~Graf and F.~D.~Steffen,
  %``Dark radiation and dark matter in supersymmetric axion models with high reheating temperature,''
  arXiv:1302.2143 [hep-ph].
  %%CITATION = ARXIV:1302.2143;%%
  %7 citations counted in INSPIRE as of 21 Jun 2013
  
  
\bibitem{Conlon:2013isa} 
  J.~P.~Conlon and M.~C.~D.~Marsh,
  %``The Cosmophenomenology of Axionic Dark Radiation,''
  arXiv:1304.1804 [hep-ph].
  %%CITATION = ARXIV:1304.1804;%%
  %4 citations counted in INSPIRE as of 21 Jun 2013


\bibitem{Conlon:2013txa} 
  J.~P.~Conlon and M.~C.~D.~Marsh,
  %``Searching for a 0.1-1 keV Cosmic Axion Background,''
  arXiv:1305.3603 [astro-ph.CO].
  %%CITATION = ARXIV:1305.3603;%%
  %1 citations counted in INSPIRE as of 21 Jun 2013


  
  \bibitem{Kelso:2013paa} 
  C.~Kelso, S.~Profumo and F.~S.~Queiroz,
  %``Non-thermal WIMPs as "Dark Radiation" in Light of ATACAMA, SPT, WMAP9 and Planck,''
  arXiv:1304.5243 [hep-ph].
  %%CITATION = ARXIV:1304.5243;%%
  %3 citations counted in INSPIRE as of 21 Jun 2013
  
  \bibitem{Angus:2013zfa} 
  S.~Angus, J.~P.~Conlon, U.~Haisch and A.~J.~Powell,
  %``Loop corrections to Delta N_eff in large volume models,''
  arXiv:1305.4128 [hep-ph].
  %%CITATION = ARXIV:1305.4128;%%
  %2 citations counted in INSPIRE as of 25 Jun 2013
  

%22 review of dark radiation from particle decay
\bibitem{Hasenkamp:2012ii} 
  J.~Hasenkamp and J.~Kersten,
  %``Dark radiation from particle decay: cosmological constraints and opportunities,''
  arXiv:1212.4160 [hep-ph].
  %%CITATION = ARXIV:1212.4160;%%
  %12 citations counted in INSPIRE as of 21 Jun 2013

%explicit models for moduli decays to axions: 9, 15, 20, 21



\bibitem{Moroi:1999zb} 
  T.~Moroi and L.~Randall,
  %``Wino cold dark matter from anomaly mediated SUSY breaking,''
  Nucl.\ Phys.\ B {\bf 570}, 455 (2000)
  [hep-ph/9906527].
  %%CITATION = HEP-PH/9906527;%%
  %352 citations counted in INSPIRE as of 21 Jun 2013


%moduli mass of 10^6-10^7 string scenarios 29-31
\bibitem{Choi:2005ge} 
  K.~Choi, A.~Falkowski, H.~P.~Nilles and M.~Olechowski,
  %``Soft supersymmetry breaking in KKLT flux compactification,''
  Nucl.\ Phys.\ B {\bf 718}, 113 (2005)
  [hep-th/0503216].
  %%CITATION = HEP-TH/0503216;%%
  %302 citations counted in INSPIRE as of 21 Jun 2013

\bibitem{Acharya:2008bk} 
  B.~S.~Acharya, P.~Kumar, K.~Bobkov, G.~Kane, J.~Shao and S.~Watson,
  %``Non-thermal Dark Matter and the Moduli Problem in String Frameworks,''
  JHEP {\bf 0806}, 064 (2008)
  [arXiv:0804.0863 [hep-ph]].
  %%CITATION = ARXIV:0804.0863;%%
  %83 citations counted in INSPIRE as of 21 Jun 2013

\bibitem{Blumenhagen:2009gk} 
  R.~Blumenhagen, J.~P.~Conlon, S.~Krippendorf, S.~Moster and F.~Quevedo,
  %``SUSY Breaking in Local String/F-Theory Models,''
  JHEP {\bf 0909}, 007 (2009)
  [arXiv:0906.3297 [hep-th]].
  %%CITATION = ARXIV:0906.3297;%%
  %79 citations counted in INSPIRE as of 21 Jun 2013


%% Moduli >10 TeV (cosmological moduli problem)


  \bibitem{de Carlos:1993jw} 
  B.~de Carlos, J.~A.~Casas, F.~Quevedo and E.~Roulet,
  %``Model independent properties and cosmological implications of the dilaton and moduli sectors of 4-d strings,''
  Phys.\ Lett.\ B {\bf 318}, 447 (1993)
  [hep-ph/9308325].
  %%CITATION = HEP-PH/9308325;%%
  %320 citations counted in INSPIRE as of 21 Jun 2013


\bibitem{Coughlan:1983ci} 
  G.~D.~Coughlan, W.~Fischler, E.~W.~Kolb, S.~Raby and G.~G.~Ross,
  %``Cosmological Problems for the Polonyi Potential,''
  Phys.\ Lett.\ B {\bf 131}, 59 (1983).
  %%CITATION = PHLTA,B131,59;%%
  %404 citations counted in INSPIRE as of 21 Jun 2013
  
  \bibitem{Ellis:1986zt} 
  J.~R.~Ellis, D.~V.~Nanopoulos and M.~Quiros,
  %``On the Axion, Dilaton, Polonyi, Gravitino and Shadow Matter Problems in Supergravity and Superstring Models,''
  Phys.\ Lett.\ B {\bf 174}, 176 (1986).
  %%CITATION = PHLTA,B174,176;%%
  %171 citations counted in INSPIRE as of 21 Jun 2013
    
  \bibitem{Banks:1993en} 
  T.~Banks, D.~B.~Kaplan and A.~E.~Nelson,
  %``Cosmological implications of dynamical supersymmetry breaking,''
  Phys.\ Rev.\ D {\bf 49}, 779 (1994)
  [hep-ph/9308292].
  %%CITATION = HEP-PH/9308292;%%
  %354 citations counted in INSPIRE as of 21 Jun 2013

%%%
  
  


%%%%%%%%%%%%

\bibitem{Kamionkowski:1990ni} 
  M.~Kamionkowski and M.~S.~Turner,
  %``Thermal Relics: Do We Know Their Abundances?,''
  Phys.\ Rev.\ D {\bf 42}, 3310 (1990).
  %%CITATION = PHRVA,D42,3310;%%
  %110 citations counted in INSPIRE as of 23 Jun 2013

\bibitem{Higaki:2013lra} 
  T.~Higaki, K.~Nakayama and F.~Takahashi,
  %``Moduli-Induced Axion Problem,''
  arXiv:1304.7987 [hep-ph].
  %%CITATION = ARXIV:1304.7987;%%
  %5 citations counted in INSPIRE as of 21 Jun 2013



\bibitem{Chung:1998rq} 
  D.~J.~H.~Chung, E.~W.~Kolb and A.~Riotto,
  %``Production of massive particles during reheating,''
  Phys.\ Rev.\ D {\bf 60}, 063504 (1999)
  [hep-ph/9809453].
  %%CITATION = HEP-PH/9809453;%%
  %160 citations counted in INSPIRE as of 23 Jun 2013

\bibitem{Fornengo:2002db} 
  N.~Fornengo, A.~Riotto and S.~Scopel,
  %``Supersymmetric dark matter and the reheating temperature of the universe,''
  Phys.\ Rev.\ D {\bf 67}, 023514 (2003)
  [hep-ph/0208072].
  %%CITATION = HEP-PH/0208072;%%
  %36 citations counted in INSPIRE as of 23 Jun 2013

\bibitem{Pallis:2004yy} 
  C.~Pallis,
  %``Massive particle decay and cold dark matter abundance,''
  Astropart.\ Phys.\  {\bf 21}, 689 (2004)
  [hep-ph/0402033].
  %%CITATION = HEP-PH/0402033;%%
  %47 citations counted in INSPIRE as of 23 Jun 2013


\bibitem{Gelmini:2006pw} 
  G.~B.~Gelmini and P.~Gondolo,
  %``Neutralino with the right cold dark matter abundance in (almost) any supersymmetric model,''
  Phys.\ Rev.\ D {\bf 74}, 023510 (2006)
  [hep-ph/0602230].
  %%CITATION = HEP-PH/0602230;%%
  %112 citations counted in INSPIRE as of 23 Jun 2013

\bibitem{Gelmini:2006pq} 
  G.~Gelmini, P.~Gondolo, A.~Soldatenko and C.~E.~Yaguna,
  %``The Effect of a late decaying scalar on the neutralino relic density,''
  Phys.\ Rev.\ D {\bf 74}, 083514 (2006)
  [hep-ph/0605016].
  %%CITATION = HEP-PH/0605016;%%
  %76 citations counted in INSPIRE as of 23 Jun 2013

%%

\bibitem{Allahverdi:2013tca} 
  R.~Allahverdi, B.~Dutta, R.~N.~Mohapatra and K.~Sinha,
  %``A Supersymmetric Model for Dark Matter and Baryogenesis Motivated by the Recent CDMS Result,''
  arXiv:1305.0287 [hep-ph].
  %%CITATION = ARXIV:1305.0287;%%
  %1 citations counted in INSPIRE as of 25 Jun 2013


\bibitem{Watson:2009hw} 
  S.~Watson,
  %``Reevaluating the Cosmological Origin of Dark Matter,''
  In *Kane, G.L. (ed.): Perspectives on supersymmetry II* 305-324
  [arXiv:0912.3003 [hep-th]].
  %%CITATION = ARXIV:0912.3003;%%
  %8 citations counted in INSPIRE as of 21 Jun 2013



\bibitem{squarksgluinos}
For an up-to-date summary, see:\\ 
\url{https://twiki.cern.ch/twiki/bin/view/AtlasPublic/SupersymmetryPublicResults} and \url{https://twiki.cern.ch/twiki/bin/view/CMSPublic/PhysicsResultsSUS}


 
  
\bibitem{micromegas}
For these calculations, we have used the publicly available tool, micrOEGAS:
  G.~Belanger, F.~Boudjema, A.~Pukhov, A.~Semenov,
  %``micrOMEGAs: A Tool for dark matter studies,''
  [arXiv:1005.4133 [hep-ph]].



\bibitem{xenon}
  E.~Aprile {\it et al.}  [XENON100 Collaboration],
  %``Dark Matter Results from 225 Live Days of XENON100 Data,''
  Phys.\ Rev.\ Lett.\  {\bf 109}, 181301 (2012)
  [arXiv:1207.5988 [astro-ph.CO]].
  %%CITATION = ARXIV:1207.5988;%%
  %149 citations counted in INSPIRE as of 24 Feb 2013






\bibitem{Griest:1990kh}
  K. Griest and D. Seckel,
  %``Three exceptions in the calculation of relic abundances,''
  Phys. Rev.  D {\bf 43}, 3191 (1991).
  %%CITATION = PHRVA,D43,3191;%%

\bibitem{Hooper:2013qjx} 
  D.~Hooper, C.~Kelso, P.~Sandick and W.~Xue,
  %``Closing Supersymmetric Resonance Regions With Direct Detection Experiments,''
  arXiv:1304.2417 [hep-ph].
  %%CITATION = ARXIV:1304.2417;%%
  %1 citations counted in INSPIRE as of 21 Jun 2013


\bibitem{Han:2013gba} 
  T.~Han, Z.~Liu and A.~Natarajan,
  %``Dark Matter and Higgs Bosons in the MSSM,''
  arXiv:1303.3040 [hep-ph].
  %%CITATION = ARXIV:1303.3040;%%

\bibitem{Beltran:2008xg} 
For a model independent treatment, see:
  M.~Beltran, D.~Hooper, E.~W.~Kolb and Z.~C.~Krusberg,
  %``Deducing the nature of dark matter from direct and indirect detection
  Phys.\ Rev.\ D {\bf 80}, 043509 (2009)
  [arXiv:0808.3384 [hep-ph]].
  %%CITATION = ARXIV:0808.3384;%%
  %61 citations counted in INSPIRE as of 05 Mar 2013

%%%%%%%%%%%%%
  
   
\bibitem{Buchmueller:2012hv} 
  O.~Buchmueller, R.~Cavanaugh, M.~Citron, A.~De Roeck, M.~J.~Dolan, J.~R.~Ellis, H.~Flacher and S.~Heinemeyer {\it et al.},
  %``The CMSSM and NUHM1 in Light of 7 TeV LHC, B_s to mu+mu- and XENON100 Data,''
  Eur.\ Phys.\ J.\ C {\bf 72}, 2243 (2012)
  [arXiv:1207.7315 [hep-ph]].
  %%CITATION = ARXIV:1207.7315;%%
  %37 citations counted in INSPIRE as of 26 Feb 2013

\bibitem{ArkaniHamed:2006mb} 
  N.~Arkani-Hamed, A.~Delgado and G.~F.~Giudice,
  %``The Well-tempered neutralino,''
  Nucl.\ Phys.\ B {\bf 741}, 108 (2006)
  [hep-ph/0601041];
  %%CITATION = HEP-PH/0601041;%%
  %142 citations counted in INSPIRE as of 07 Mar 2013
  M.~Farina, M.~Kadastik, D.~Pappadopulo, J.~Pata, M.~Raidal and A.~Strumia,
  %``Implications of XENON100 and LHC results for Dark Matter models,''
  Nucl.\ Phys.\ B {\bf 853}, 607 (2011)
  [arXiv:1104.3572 [hep-ph]];
  %%CITATION = ARXIV:1104.3572;%%
  %64 citations counted in INSPIRE as of 07 Mar 2013
  C.~Cheung, L.~J.~Hall, D.~Pinner and J.~T.~Ruderman,
  %``Prospects and Blind Spots for Neutralino Dark Matter,''
  arXiv:1211.4873 [hep-ph].
  %%CITATION = ARXIV:1211.4873;%%
  %6 citations counted in INSPIRE as of 21 Mar 2013



\bibitem{Feng:2011aa} 
  J.~L.~Feng, K.~T.~Matchev and D.~Sanford,
  %``Focus Point Supersymmetry Redux,''
  Phys.\ Rev.\ D {\bf 85}, 075007 (2012)
  [arXiv:1112.3021 [hep-ph]].
  %%CITATION = ARXIV:1112.3021;%%
  %58 citations counted in INSPIRE as of 26 Feb 2013



\bibitem{lux} 
  D.~S.~Akerib {\it et al.}  [LUX Collaboration],
  %``Technical Results from the Surface Run of the LUX Dark Matter Experiment,''
  arXiv:1210.4569 [astro-ph.IM].
  %%CITATION = ARXIV:1210.4569;%%
  %5 citations counted in INSPIRE as of 27 Feb 2013


\bibitem{xenon1t} 
  E.~Aprile [XENON1T Collaboration],
  %``The XENON1T Dark Matter Search Experiment,''
  arXiv:1206.6288 [astro-ph.IM].
  %%CITATION = ARXIV:1206.6288;%%
  %20 citations counted in INSPIRE as of 27 Feb 2013



\bibitem{Steigman:2012nb} 
  G.~Steigman, B.~Dasgupta and J.~F.~Beacom,
  %``Precise Relic WIMP Abundance and its Impact on Searches for Dark Matter Annihilation,''
  Phys.\ Rev.\ D {\bf 86}, 023506 (2012)
  [arXiv:1204.3622 [hep-ph]].
  %%CITATION = ARXIV:1204.3622;%%
  %32 citations counted in INSPIRE as of 21 Jun 2013

  
\bibitem{GeringerSameth:2011iw} 
  A.~Geringer-Sameth and S.~M.~Koushiappas,
  %``Exclusion of canonical WIMPs by the joint analysis of Milky Way dwarfs with Fermi,''
  Phys.\ Rev.\ Lett.\  {\bf 107}, 241303 (2011)
  [arXiv:1108.2914 [astro-ph.CO]].
  %%CITATION = ARXIV:1108.2914;%%
  %98 citations counted in INSPIRE as of 21 Jun 2013  
  
\bibitem{Ackermann:2011wa} 
  M.~Ackermann {\it et al.}  [Fermi-LAT Collaboration],
  %``Constraining Dark Matter Models from a Combined Analysis of Milky Way Satellites with the Fermi Large Area Telescope,''
  Phys.\ Rev.\ Lett.\  {\bf 107}, 241302 (2011)
  [arXiv:1108.3546 [astro-ph.HE]].
  %%CITATION = ARXIV:1108.3546;%%
  %204 citations counted in INSPIRE as of 21 Jun 2013  


\bibitem{Hooper:2012sr} 
  D.~Hooper, C.~Kelso and F.~S.~Queiroz,
  %``Stringent and Robust Constraints on the Dark Matter Annihilation Cross Section From the Region of the Galactic Center,''
  arXiv:1209.3015 [astro-ph.HE].
  %%CITATION = ARXIV:1209.3015;%%
  %21 citations counted in INSPIRE as of 21 Jun 2013


  
  
  

%% Moduli

\bibitem{Dine:1995kz} 
  M.~Dine, L.~Randall and S.~D.~Thomas,
  %``Baryogenesis from flat directions of the supersymmetric standard model,''
  Nucl.\ Phys.\ B {\bf 458}, 291 (1996)
  [hep-ph/9507453].
  %%CITATION = HEP-PH/9507453;%%
  %455 citations counted in INSPIRE as of 21 Jun 2013





%%% not yet used

%\bibitem{Weinberg:2013kea} 
 % S.~Weinberg,
 % %``Goldstone Bosons as Fractional Cosmic Neutrinos,''
 % Phys.\  Rev.\  Lett.\  110, {\bf 241301} (2013)
%  [arXiv:1305.1971 [astro-ph.CO]].
  %%CITATION = ARXIV:1305.1971;%%
  %5 citations counted in INSPIRE as of 21 Jun 2013



%%%%%%%%%%%%%%%%



%%%%%%%%%%%%%%%%%%%%%%


%%%%%%%%%%%%%%%%%%%

\end{thebibliography}
\end{document}